\documentclass{article}
\usepackage{spconf,amsmath,graphicx}
\usepackage[ruled,linesnumbered]{algorithm2e}
\usepackage{algpseudocode}
\usepackage{multirow}
\usepackage{balance}
\usepackage{fancyhdr}
\usepackage{hyperref}
\usepackage{cleveref}

\title{ConchShell: A Generative Adversarial Networks that Turns Pictures into Piano Music}
\name{Wanpeng Fan$^{\dagger}$ \qquad Yuanzhi Su$^{\dagger}$ \qquad Yuxin Huang$^{\star}$}
  
\address{$^{\dagger}$ School of Computer Science and Cyber Engineering, Guangzhou University, Guangzhou, China\\
 $^{\star}$ Law School, Guangzhou University, Guangzhou, China}
\begin{document}
%
\maketitle
\begin{abstract}
We present ConchShell, a multi-modal generative adversarial framework that 
takes pictures as input to the network and generates piano music samples that 
match the picture context.Inspired by I3D, we introduce a novel image feature 
representation method: time-convolutional neural network (TCNN), which is used
to forge features for images in the temporal dimension. Although our image data
consists of only six categories, our proposed framework will be innovative and 
commercially meaningful. The project will provide technical ideas for work such as
3D game voice overs, short-video soundtracks, and real-time generation of 
metaverse background music.We have also released a new dataset, 
the Beach-Ocean-Piano(BOP) dataset 
\footnote{Hyperlink to BOP dataset: \url{https://drive.google.com/drive/folders/1iTHo76nJzvPQ6_GIQbokfdRvy6pcN9ti}\label{bop}},
which contains more than 3,000 images and more than 1,500 piano pieces.
This dataset will support multimodal image-to-music research.
\end{abstract}

\section{Introduction}
\label{sec:intro}
Music is the art of time. It appeals to people's auditory senses and expresses people's emotions
through sound. At the same time, painting is an art of space, appealing to people's visual
enses through lines and colors, and expressing the stories in people's minds.
In an artistic activity, the mutual sense of hearing and vision is the most active and influential.
Based on this idea, we propose a model, ConchShell, to connect music and images, and
it will be a project full of artistic significance, commercial significance and scientific significance.

In the field of art, many scholars are interested in the art composed of music and 
painting together\cite{Bogue2003Deleuze,bogue2014deleuze}, and some of them discuss the 
relationship between music and painting from the perspective of history and 
theory\cite{morton2013arts,shevchuk2018interspecific}. In ancient China, there was also the 
habit of poets composing poems for ink painting\cite{ferreira2021painting}. 
Therefore, studying the association between music signals and picture signals essentially 
provides a theoretical basis for the integration and progress of art.

\begin{figure}
  \centering 
  \includegraphics[width=85mm]{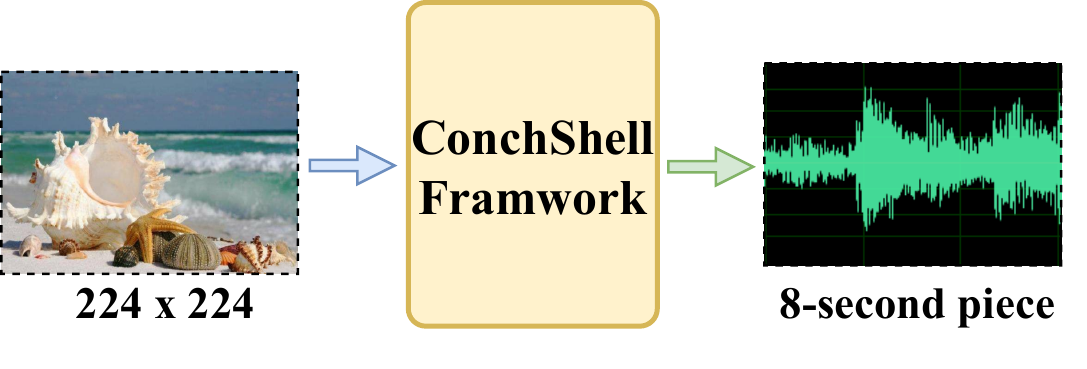}
  \caption{Working diagram of ConchShell.  The size of the input picture is 224x224. 
  After encoding and decoding by ConchShell, a single channel 
  piano piece with a length of 8 seconds will be generated.}
  \label{conch_shell1}
\end{figure}

Meanwhile, excellent 3D games and Open-World games have become a way
for people to entertain.
Players roam in the virtual world, let their mind shocked,
in addition to the fantastic images, and surround their ears with music.
Undeniably, the exquisite pictures are an important factor for 
a game to arouse the enthusiasm of players. However, the plot and music are 
the key factors to determine whether the players can be retained. 
However, composing music for virtual worlds is very expensive, such as Genshin Impact and Dungeon\&Fighter,
and composing music for movies can cost millions of dollars or more,
Therefore, our project will create great commercial value and reduce the expenditure on soundtrack of excellent works.

In order to realize the mapping of pictures to music, this paper proposes the following contributions:
\begin{itemize}
\item We propose ConchShell, a multimodal GAN framework that will take an image and generate an 8-second piano piece (Figure \ref{conch_shell1}).
\item This model uses TCNN, which uses fewer neurons than VGG \cite{simonyan2014very} 
to forge a third dimension for the picture: time dimension, 
which makes the connection between the picture and the music more explicit.
\item We have created BOP dataset, which contains manual annotation 
of image-music labels, and we will release this dataset in the future to provide 
data support for multi-modal academic research on images and music for free.

\end{itemize}

\section{Related Work}
\label{sec:relatedWork}
We did not find a reliable open source project 
before attempting the task of converting images to music. 
For this reason, we turned our goal to similar multimodal tasks in an attempt to find common ground on these studies.
\subsection{Music Generation}
\label{ssec:musicGeneration}
Due to the high dimensionality of audio data and its complex temporal correlation, 
the computational overhead is unimaginable. 
As a result, many models use intermediate audio representations to learn generative models 
in order to reduce the computational requirements.

MelNet \cite{vasquez2019melnet} uses a two-dimensional frequency domain signal converted to a time domain signal, 
while MidiNet \cite{yang2017midinet} is a CNN-based GAN framework that generates MIDI notes one after the other in the symbolic domain. 
Our ConchShell framework, on the other hand, uses the VQ representation of Jukebox \cite{dhariwal2020jukebox}, 
a model for generating music in the raw audio domain. 
It uses multi-scale VQ-VAE to compress raw audio into discrete codes and models them using autoregressive Transformers.
\subsection{Images, Videos and Music}
\label{ssec:image_et_al}
The biggest difference between pictures and music is that music has a complex temporal correlation, 
but pictures do not. This means that pictures may not contain information that can be mapped to the melody of the music. 
Video, however, has temporal correlation. In Two-Stream Inflated 3D ConvNet (I3D) \cite{carreira2017quo}, 
the model learns seamless spatio-temporal features from video, 
and knowing that video is actually an ordered collection of multiple pictures, 
then attempting to associate pictures with video, i.e. pre-converting pictures to have some form of temporal features, 
can achieve the goal of converting pictures to music.
\section{Our Method}
\label{sec:ourMethod}
\begin{figure}
  \begin{center}
    \includegraphics[width=86mm]{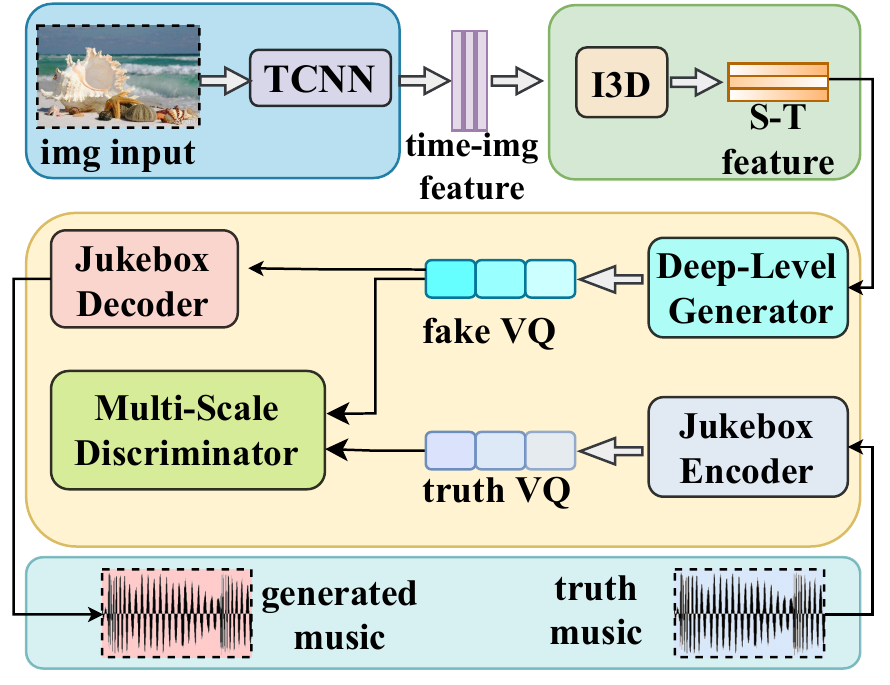}
  \end{center}
  \caption{ConchShell-Framework. Details of how this framework works.}
  \label{conch_shell2}
\end{figure}

The process of converting images to music can be briefly summarised as follows: 
1) extract enough features from the images. 
2) compress these features to get some sort of underlying musical information.
3) map this information to musical wave.
The overall model is illustrated in Fig\ref{conch_shell2}.
\subsection{Time-Convolutional Neural Network}
\label{ssec:tcnn}
To extract features from the images, we designed the Time-Convolutional Neural Network (TCNN), 
which is a classification network consisting of 6 convolutional blocks combined. 
After training, we will extract features from the convolutional blocks, 
and each layer of features will be repeatedly spliced 3 times to fake the temporal dimension for the images, 
the final features will contain 3 image channels and 18 temporal channels, the specific pseudo-code is as follows:

\begin{algorithm}
\label{algorithm}
\caption{Get Time-Image-Feature}
\KwData{A list of image feature $f_{image}$}
\KwResult{Image features containing temporal information $f_{t-img}$}
$f^{*}_{t-img}\leftarrow []$\;
\For{$i\leftarrow 0$ \KwTo 5 }{
  $f^{*} \leftarrow mean(f_{image}[i],dim=0)$;
  
  \Comment{[3, 1, 224, 224]}

  \For{$j\leftarrow 0$ \KwTo 2}{
    $f^{*}_{t-img}$.append($f^{*}$);
  }
}
$f_{t-img} \leftarrow$cat($f^{*}_{t-img}$,dim=1);\Comment{[3, 18, 224, 224]}
\end{algorithm}
Space is limited so only a simplified version of the pseudo-code is written here, 
but rest assured that we will open source the entire code for this project.

In order to verify the effectiveness of the method, we also used VGG networks \cite{simonyan2014very} for feature extraction 
and detailed comparison results can be found in \cref{tab1,tab2}.
\subsection{I3D and Jukebox}
\label{ssec:i3d_and_Jukebox}
The T-Img-Features of the TCNN will be used as input to I3D \cite{carreira2017quo}, 
a pre-trained model, and output spatio-temporal features (S-T-Features) 
which will be fed into the ConchShell generator.

Jukebox is also a pre-trained model that contains High, Low and Deep levels of decoder. 
The High-Level model will focus more on the melody and temporal relevance of the generated music,
but will ignore the quality of the generated audio. The Low-Level and Deep-Level models 
generate audio with less noise, but their ability to focus on the overall melody is diminished. 
We have designed three different generators for these three levels to take the best out of the worst as far as possible.

\subsection{Generators and Multi-Scale Discriminator}
\label{ssec:generatorDiscriminator}
Inspired by Zhu, Ye et al \cite{zhu2022quantized}, we follow the Multi-Scale Discriminator they designed, 
the discriminator that has worked in many audio synthesis tasks in recent years \cite{kong2020hifi,kumar2019melgan}.
In our project, the discriminator will accept two inputs, Vector Quantized ($VQ_{1}$) \cite{van2017neural}generated
from ConchShell-Generator and ground truth $VQ_{2}$ from Jukebox-Encoder.

We have modified the generator, which is essentially a convolutional network with a kernel size of up to 40, 
in order for it to capture melodically relevant information, 
and multiple residual structures between each convolutional layer to increase the perceptual field 
as well as to adjust its details, finally passing through a linear layer with 16k output channels, 
which is designed to match the output format of Jukebox.

\subsection{Loss Function}
\label{ssec:loss}
We have adopted the following loss function, drawing on the approach of Zhu, Ye et al \cite{zhu2022quantized}.

\textbf{Discriminator Loss.}The loss function used by the discriminator is essentially a hinge loss \cite{miyato2018spectral}.
\begin{equation}
  \begin{aligned}
      L_D(G,D) &= \sum_{k=1}^{3}{E[(\max(0,1-D_k(J_{ukebox}^{en}(x_g)))]}\\
                &+ \sum_{k=1}^{3}{E[\max(0,1+D_k(G(x_{img})))]}
  \end{aligned}
\end{equation}   
$x_{g}$ represents the original audio which will be used as input to Jukebox's encoder. 
$x_{img}$ is a image which will be fed into ConchShell.
k is the number of layers in the discriminator and we have empirically set k to 3.

\textbf{Generator Loss} is a modified version of hinge loss.
\begin{equation}
  \begin{aligned}
      L_G(G,D) = \sum_{k=1}^{3}{E[(-D_k(G(x_{img}))]}
  \end{aligned}
\end{equation}   

\textbf{Feature Matching Loss.} The feature matching loss \cite{larsen2016autoencoding} is defined as the L1 loss, 
whose inputs are the real VQ features and the generated VQ features, 
and this loss is used to encourage the adjustment of some detail information in the audio signal.
\begin{equation}
  \begin{aligned}
    L_{FM}&(G,D) = \\
    &\frac{1}{N_i}\sum_{i=1}^{T}{\left\|(D^i(J_{ukebox}^{en}(x_g))-D^i(G(x_{img}))\right\|}
  \end{aligned}
\end{equation}

\textbf{Audio Perceptual Losses.} The perceptual loss is defined as 
the L1 distance between the original audio and the generated audio samples.
We expect the perceptual loss of audio signals from the time and Mel domains to further improve the quality of listening.
\begin{equation}
  \begin{aligned}
      L_{wave}(G) = L_{1}(x_{g},J_{ukebox}^{de}(G(x_{img})))
  \end{aligned}
\end{equation}
\begin{equation}
  \begin{aligned}
    L_{Mel}(G) = L_{1}(\varphi (x_{g}),\varphi (J_{ukebox}^{de}(G(x_{img}))))
  \end{aligned}
\end{equation}
where $\phi$  is the function to compute the mel-spectrogram features for the audio signal waveforms.
The generated audio will be used as input to Jukebox's decoder. 

\textbf{ConchShell Loss.} The final training objective for our framework is defined as follows:
\begin{equation}
  \begin{aligned}
    L_{CS} = \lambda_{g} L_{G}+\lambda_{fm} L_{FM}+\lambda_{w} L_{wave}+\lambda_{m} L_{Mel}
  \end{aligned}
\end{equation}
where the $\lambda_{g}$, $\lambda_{fm}$, $\lambda_{w}$, and $\lambda_{m}$ are hyperparameters,
and in our experiments they were adjusted to 0.5, 3, 40 and 15.
\section{Experiments}
\label{sec:experiments}
\subsection{Datasets}
\label{subsec:datasets}
We created Beach-Ocean-Piano Datasets (BOP), a dataset containing six image categories with approximately 3k images. 
We split it into a training set and a validation set, 
with the training set accounting for $85\%$ of the total number of images. 
The images were manually annotated with audio that matched the context of the images, 
the emotion of the music and the background story.

The audio was all piano-only, single-channel, and resampled to 16k,
and we decided to set the duration of the music to 8 seconds, 
considering that too short a duration would make it difficult to convey the emotion and 
too long would increase the computational overhead. The size of the images was limited to 224x224, 
and in order to retain the emotional information attached to the colours, 
we used coloured three-channel images.
\begin{table*}[h!]
  \begin{center}
    
    \begin{tabular}{cccccccccc} 
      \hline
      \textbf{GAN-Model} & \multicolumn{3}{c}{\textbf{D-Module}} &\multicolumn{3}{c}{\textbf{G-Module}} &\multicolumn{3}{c}{\textbf{Img-Module}}\\
      \hline 
      &Model&FLOPs&Params&Model&FLOPs&Params&Model&FLOPs&Params\\
      \hline
      \textbf{Baseline-1}&D-High&2.734G&9.212M&G-High&23.800G&33.109M&TCNN-6&827.507M&2.285M\\
      \hline
      \textbf{Baseline-2}&D-Low&10.976G&9.212M&G-Low&27.205G&65.183M&TCNN-6&827.507M&2.285M\\
      \hline
      \textbf{ConchShell}&D-Deep&43.950G&9.212M&G-Deep&59.197G&100.420M&TCNN-6&827.507M&2.285M\\
      \hline
      \textbf{Baseline-3}&D-Deep&43.950G&9.212M&G-Deep&59.197G&100.420M&TCNN-18&1.443G&7.543M\\
      \hline
    \end{tabular}
    \caption{A comparison of the composition of ConchShell with each baseline model.}
    \label{tab1}
  \end{center}
\end{table*}

\begin{table*}[h!]
  \begin{center}
    
    \begin{tabular}{ccccccc}
      \hline
      \textbf{Data-Type}&\textbf{GAN-Model} & \textbf{PESQ} &\textbf{STOI} &\textbf{I2M-MOS}&\textbf{Muse-MOS}&\textbf{Clean-MOS}\\
      \hline

      \multirow{4}{*}{train-set}&\textbf{Baseline-1}&1.4&0.32&1.7 $\pm$ 0.4&1.6 $\pm$ 0.3&1.7 $\pm$ 0.3\\

      &\textbf{Baseline-2}&1.8&0.55&2.7 $\pm$ 0.4&2.6 $\pm$ 0.5&2.3 $\pm$ 0.1\\
      
      &\textbf{ConchShell}&\textbf{3.1}&\textbf{0.73}&\textbf{3.5 $\pm$ 0.4}&\textbf{3.3 $\pm$ 0.5}&3.7 $\pm$ 0.2\\
      
      &\textbf{Baseline-3}&2.7&0.69&2.9 $\pm$ 0.3&2.8 $\pm$ 0.3&\textbf{4.1 $\pm$ 0.2}\\
      \hline
      \multirow{4}{*}{test-set}&\textbf{Baseline-1}&1.4&0.33&1.5 $\pm$ 0.5&1.5 $\pm$ 0.3&1.5 $\pm$ 0.2\\

      &\textbf{Baseline-2}&1.5&0.49&2.5 $\pm$ 0.7&2.4 $\pm$ 1.0&2.4 $\pm$ 0.2\\
      
      &\textbf{ConchShell}&\textbf{2.9}&0.66&\textbf{2.8 $\pm$ 0.5}&\textbf{3.0 $\pm$ 0.6}&3.7 $\pm$ 0.2\\
      
      &\textbf{Baseline-3}&2.7&\textbf{0.67}&2.4 $\pm$ 0.4&2.5 $\pm$ 0.4&\textbf{4.0 $\pm$ 0.3}\\
      \hline
      truth-data&-&-&-&4.6 $\pm$ 0.3&5.0&5.0\\
      \hline
    \end{tabular}
    \caption{The evaluation results for each model.}
    \label{tab2}
  \end{center}
\end{table*}

\subsection{Training Details}
\label{subsec:training}
We used a multi-stage training strategy, and TCNN was trained using this dataset, 
whose goal was to classify six images. Whereas ConchShell is difficult to train, 
we use a distributed training approach on four Tesla-4 GPUs in distributed data parallel (DDP) mode, 
and because its model is large, the batch size we set to 2.

Both the generator and discriminator use AdamW as the optimizer, $\beta_1=0.5$ and $\beta_2=0.999$. 
Their learning rates are initialized to 1e-5, with the generator's learning rate reduced to 0.8 times 
and the discriminator reduced to 0.9 times for every 20k steps run.

We also trained three other baseline models with the same training strategy, 
except for a slightly different batch size. The exact combination of 
baseline models can be found in \cref{tab1}.

\subsection{Evaluations}
\label{subsec:eval}
After training, we chose two objective metrics to evaluate the quality of the generated music, 
namely Perceptual Evaluation of Speech Quality (PESQ) \cite{rix2001perceptual} and Short-Time Objective Intelligibility (STOI), 
while we drew on Mean opinion scores (MOS) and designed several subjective evaluation metrics, 
namely Image-to-Muse-MOS (I2M), Melody-MOS, and Noise-MOS.

The evaluators will evaluate the generated audio using the above three subjective metrics, 
where I2M-MOS represents the relevance of the image to the music, with a score range of 1-5, 
where 1 means the generated music is not relevant to the image at all, 
and 5 means the generated music perfectly expresses the connotations of the image.
Melody-MOS refers to the generated music itself and is used to assess 
whether the generated music has a melodic structure that is acceptable to humans, 
i.e. whether the generated music is temporally relevant, where 1 means that 
the generated music is completely incompatible with human auditory aesthetics, 
e.g. the generated music repeats a single note all the time or 
the generated notes do not have any pattern, and 5 means that the generated music has a strong melodic structure.
Finally, the Noise-MOS measures how much background noise the generated music contains, 
with 1 meaning that the noise almost covers the original music, 
and 5 meaning that no noise is heard at all.

During this process, the evaluators were asked to repeat the evaluation twice and 
we averaged the results from these two occasions. This resulted in \cref{tab2}. 
From the results we can see that our ConchShell outperformed the baseline model in almost all metrics.
As can be seen from the results of Baseline-3, using only VGG for feature extraction on images 
without falsifying the temporal dimension leads to a decrease in the values of I2M-MOS and Melody-MOS. 
We believe that I3D is insensitive to irregular features, 
resulting in little to no mapping of the S-T-Features it generates to music.
At the same time, we observed that the values in the test set were lower than 
those in the training set, probably because we had less training data, 
which made the model over-fit, and increasing the amount of data should improve this situation.
We can then see that, regardless of the model, the Noise-MOS results do not reach the desired values, 
which means that the music generated is always noisy, 
so it is better to perform noise reduction \cite{porov2018music} for all results if you want to get high quality music.

\section{Conclusion}
\label{sec:conclusion}
We have released a new dataset, Beach-Ocean-Piano(BOP) dataset, 
which will be free and open source, to provide data support for multi-modal tasks with images and music.
In the meantime we created ConchShell, a multi-modal framework based on GAN, 
which will take an image and generate an 8-second piano piece for it.
To better associate images with music, we also designed the Time-Convolutional Neural Network (TCNN), 
which will work with the I3D model to transform images into features with a temporal dimension.
This framework will provide data support and research ideas for other multimodal tasks, 
especially image-to-music tasks.
\balance
\vfill\pagebreak
\bibliographystyle{IEEEbib}
\bibliography{strings}

\end{document}